\documentclass[universe,article,accept,pdftex,moreauthors]{mdpi} 
\firstpage{1} 
\pubvolume{1}
\issuenum{1}
\articlenumber{0}
\pubyear{2022}
\copyrightyear{2022}
\datereceived{} 
\dateaccepted{} 
\datepublished{} 
\hreflink{https://doi.org/} % If needed use \linebreak

\Title{Magnetar wind-driven shock breakout emission after double neutron star mergers: The effect of the anisotropy of the merger ejecta}

\TitleCitation{Title}

\Author{Guang-Lei Wu$^{1}$, Yun-Wei Yu$^{1,2}$*\orcidA{} and Shao-Ze Li$^{3}$\orcidB{}}

\AuthorNames{Firstname Lastname, Firstname Lastname and Firstname Lastname}

\AuthorCitation{Lastname, F.; Lastname, F.; Lastname, F.}
\address{%
	$^{1}$ \quad Institute of Astrophysics, Central China Normal
	University, Wuhan 430079, China\\
	$^{2}$ \quad Key Laboratory of Quark and Lepton Physics (Central
	China Normal University), Ministry of Education, Wuhan 430079,
	China\\
	$^{3}$ \quad College of Physics Science and Technology, Hebei University, Baoding 071002,
	China}
\corres{Correspondence: yuyw@ccnu.edu.cn}

\linenumbers
\abstract{ A rapidly rotating and highly magnetized remnant neutron star (NS; magnetar) could survive from a merger of double NSs and drive a powerful relativistic wind. The early interaction of this wind with the previous merger ejecta can lead to shock breakout (SBO) emission mainly in ultraviolet and soft X-ray bands, which provides an observational signature for the existence of the remnant magnetar. Here we investigate the effect of an anisotropic structure of the merger ejecta on the SBO emission. It is found that bolometric light curve of the SBO emission can be broadened, since the SBO can occur at different times for different directions. In more detail, the profile of the SBO light curve can be highly dependent on the ejecta sturcture and, thus, we can in principle use the SBO light curves to probe the structure of the merger ejecta in future.}


\begin{document}

%%%%%%%%%%%%%%%%%%%%%%%%%%%%%%%%%%%%%%%%%%

\section{Introduction}
Kilonovae are optical transient emission originating from mergers of double neutron stars (NSs) or NS-black hole (BH) binaries \cite{Li1998,Metzger2010,Roberts2011,Barnes2013,Kasen2013,Tanaka2013}. The power source of the kilonova emission is conventionally considered to be the radioactive decays of $r$-process elements, as these elements can be synthesized effectively in the neutron-rich merger ejecta. As an electromagnetic counterpart of the merger-induced GW radiation, kilonova emission can play a very important role in confirming the position, time, redshift, and even the progenitor properties of the mergers. Such effects of kilonovae had been manifested completely in the famous GW170817 event \cite{Abbott2017b}, which was accompanied with a kilonova AT2017gfo \cite{Abbott2017c,Arcavi2017,Coulter2017,Cowperthwaite2017,Evans2017,Nicholl2017,Smartt2017,Soares-Santos2017,Tanvir2017}. The detailed modeling of the observed kilonovae and as well as their associated gamma-ray burst (GRB) afterglows could further give constraint on the properties of the merger products, especially, when the traditional kilonova model is challenged by the observations.    

The nature of the remnant of double NS mergers is one of the most concerned issues in current astrophysical studies. The answer to this mystery would help to constrain the mass limit of NSs and thus the equation of state (EOS) of nuclear matter, which is highly related to the unclear non-perturbative quantum chromodynamics at low energies. However, the confirmation of the nature of merger products is beyond the ability of the current GW detectors, since the detector sensitivities are still higher than the potential GW radiation from the merger products \cite{Abbott2017a}. Therefore, alternatively, it is expected that the electromagnetic counterparts of the GW events can provide other observational signatures for judging the nature of the merger products. Following this consideration, Yu et al. (2013) \citep{Yu2013} and Metzger \& Piro (2014) \citep{Metzger2014b} investigated the influence of a post-merger NS on the kilonova emission, where the remnant NS is considered to be rapidly rotating and highly magnetized, i.e., a millisecond magnetar. As the most direct effect, the spin-down of the millisecond magnetar can inject energy into the merger ejecta and then enhance the kilonova emission and accelerate the ejecta significantly. Thus, a new name of mergernova was suggested to replace the traditional kilonova, in order to reflect the predicted wide range of the emission luminosity that depends on the specific properties of the remnant NS \cite{Yu2013}. Using this magnetar-driven mergernova model, Yu et al. (2018) successfully accounted for the relatively high luminosity of AT 2017gfo on the order of $\sim10^{42}\rm erg~s^{-1}$ \cite{Yu2018}, whereas the radioactive kilonova model needs to invoke a too high ejecta mass 
 \cite{Li2018,Shibata2017,Metzger2018} (cf., Reference \cite{Kawaguchi2018} showed this difficulty could somewhat overcome by considering the 2D radiative transfer elaborately).

Frankly speaking, in view of the complication of mergernova emission, only the information of the luminosity and temperature of the emission is not enough for confirming the existence of a post-merger magnetar. The detailed transfer and transformation of the spin-down energy of the magnetar need to be investigated further \cite{Metzger2014b,Li2016,Gao2015,Siegel2016a,Siegel2016b,Gao2017,Yu2018,Li2018,Wollaeger2019,Yu2019,Ren2019,Ren2022,Wu2021,Ai2022}, which could lead to some extra independent observational signatures for the magnetar. Specifically, the energy release from the spinning-down magnetar could be initally in the form of a Poynting flux and gradually transform into a relativistic wind consisting of electron/positron pairs. Once this magnetar wind catches up and collides with the preceding merger ejecta, it will generate a forward shock (FS) propagating into the ejecta and a termination shock (TS) reversely into the continuously injected wind. After a short period, the FS would break out from the ejecta, while the TS is long lasting. So, the primary channel of the energy transfer is the absorption of the TS emission (i.e., a pulsar wind nebula emission) by the merger ejecta. The TS emission can finally leak from the merger ejecta as the ejecta becomes transparent in the related electromagnetic band. As a rsult, a non-thermal emission component can appear in the late phase of the mergernova emission, which have been found in many mergernova candidates including AT2017gfo \cite{Gao2015,Gao2017,Ren2019,Yu2019,Wu2021,Ren2022}. Meanwhile, the breakout of the FS could also cause a rapid soft X-ray flare prior to the primary mergernova emission \cite{Li2016}, which is similar to the situation discovered in  the supernovae that are suggested to be driven by a magnetar too \cite{Kasen2016,Liu2021,Zhang2022}. Generally, this shock breakout (SBO) precursor emission of a mergernova is likely to be outshone by more luminous afterglow emission of the associated GRB. Neverthelss, the observation of GRB 170817A indicates that the nearby GW events are very probably observed off-axis, which can significantly suppress the early afterglow emission and then make the SBO emission emerging.

Therefore, at present, it is necessary to investigate in advance more detailed features of the observational signatures of the magnetar-driven mergernovae, particularly, in expectation of the future discovery of the SBO emission. In theory, the propagation of the FS into the merger ejecta is determined by the radial structure of the merger ejecta. The description of the ejecta structure requires detailed numerical simulations. The first simulation on this topic was performed by \citeauthor{Davies1994} (\citeyear{Davies1994}) with an analytical EOS of nuclear matter and a Newtonian gravity \cite{Davies1994}, which showed that $\sim2\% $ mass of the system can be ejected around the equatorial plane as the dynamical tail of the spiral tidal arms. This result was subsequently reproduced by more simulations \cite{Ruffert1996,Rosswog1999,Rosswog2000,Ruffert2001,Korobkin2012}, which further found that (i) the neutrino-energy deposition in the cool outer regions of the 
tidal arms can enhance the mass loss and (ii) the material at the contact interface between the NSs can be squeezed out although the tidal ejecta could still be dominant.
Subsequently, the consideration of general relativity and more realistic EOSs leads to some new understandings of the formation of the ejecta \cite{Piran2013,Rosswog2013,Bauswein2013,Hotokezaka2013,Sekiguchi2016,Foucart2016,Radice2016,Dietrich2017a,Dietrich2017b,Radice2018,Shibata2021}. It is showed that the squeezed component of the ejecta could become dominant over the tidal component \cite{Oechslin2007,Bauswein2013,Hotokezaka2013}, as the squeezed material is effectively heated and accelerated by a shock, spreading into a quasi-isotropic region and owning a velocity higher than that of the tidal component. Then, the collision of the squeezed component with the tidal component would lead to a more complex structure of the ejecta. On the other hand, the simulations also showed that a thick disk of a mass of $\sim 0.1M_{\odot}$ can surround the merger remnant and the neutrino emission from both the disk and remnant can drive a hot wind into a funnel shape region in the polar direction \cite{Dessart2009,Perego2014,Martin2015,Fujibayashi2017}. The disk wind component can last for a relatively long timescale due to the viscous heating in the disk, which is further dependent on the lifetime of the remnant NS and the magnetic field configuration of the disk \cite{Siegel2014,Fujibayashi2018,Siegel2018,Fujibayashi2018,Fernandez2019}. As the accretion disk finally expands and becomes advection dominated, the wind would become more isotropic and slower until the accretion becomes inefficient \cite{Fernandez2013,Just2015,Fujibayashi2018,Fernandez2019}. In summary, the properties of the merger ejecta can vary obviously with the angle relative to the symmetric axis of the system, which may leave imprints in the SBO emission. Therefore, in this paper, we revisit the SBO emission by taking into account the anisotropic structure of the merger ejecta.

%%%%%%%%%%%%%%%%%%%%%%%%%%%%%%%%%%%%%%%%%%
\section{The model}
\subsection{Anisotropic merger ejecta}

\begin{figure*}[htpb]
	\centering
	\includegraphics[width = 0.7\linewidth ]{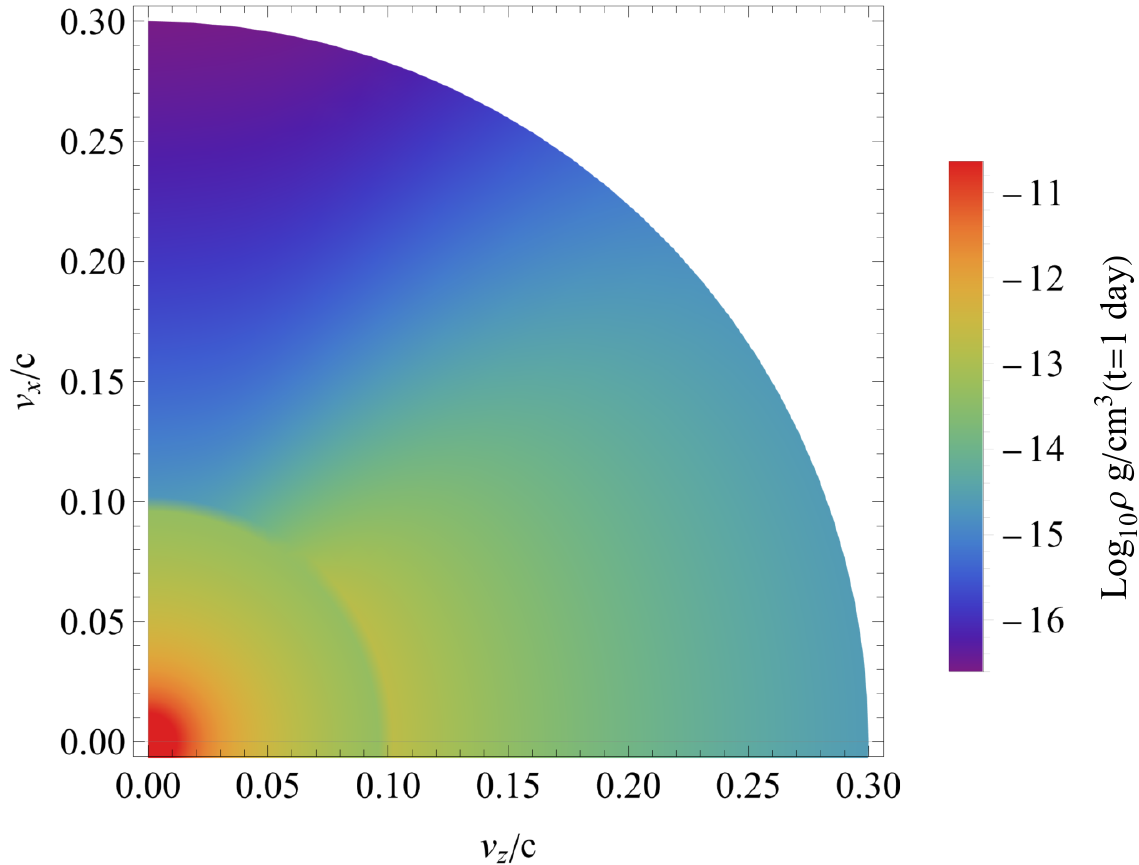}
	\caption{The density profile employed in our calculations.}
	\label{densityplot}
\end{figure*}

As introduced, the material ejected during a merger of double NSs can have different origins, which at least contain three channels including the tidal and squeezed dynamical origins and the disk wind origin. These ejecta components own different masses, velocities, electron fractions, and element compositions, which leads the merger ejecta to be highly anisotropic. In this paper, we adopt the empirical description of the ejecta structure suggested by Kawaguchi et al.  \cite{Kawaguchi2018,Kawaguchi2020}. While the disk wind is considered to be roughly isotropic, the mass distribution of the dynamical ejecta can exhibit an angular dependence as follows: \cite{Kawaguchi2018,Kawaguchi2020}
\begin{equation}
	\eta\left(\theta\right)=\left[1-\Theta\left(\theta\right)\right]f_{\mathrm{}}+\Theta(\theta),\label{angular1}
\end{equation}
where $ f_{\mathrm{}}$ represents the density ratio of $\theta=0$ to $\pi/2$ and
\begin{equation}
	\Theta\left(\theta\right)=\frac{1}{1+\mathrm{exp}\left[-10\left(\theta-\pi/4\right)\right]}.\label{angular2}
\end{equation}
 The above expressions were summarized from the numerical simulations such as those given by References \cite{Radice2016,Sekiguchi2016,Kiuchi2017,Shibata2017}, which describe the combination of the tidal and squeezed components. The propagation of the magnetar wind-driven FS in the ejecta is highly dependent on the radial distribution of the ejecta mass. According to numerical simulations, we assume as usual that the radial profile of the different ejecta components can be described by power laws as follows \cite{Nagakura2014}
\begin{equation}
	\rho_{\mathrm{ej}} \left(r ,\theta \right) \propto\begin{cases}
		r^{-3},& {\rm~for~the~wind~component,}% r_{\min,\mathrm{w}}  <r < r_{\max,\mathrm{w}},
		\\
		\eta\left(\theta\right)r^{-4},& {\rm~for~the~dynamical~component,}%   r_{\min,\mathrm{d}}  <r < r_{\max,\mathrm{d}},\\
	\end{cases}\label{rhor}
\end{equation}
which can be normalized by the masses of the disk wind ejecta $M_{\mathrm{ej,w}} $ and the dynamical ejecta $M_{\mathrm{ej,d}}$, respectively. This 2D distribution of the density of the merger ejecta in the  $v_{x}-v_{z}$ plane is shown in Fig. \ref{densityplot}, for the parameter values of Model A listed in Table \ref{table1}. An axial symmetry around the $z-$axis is adopted. Here, the maximum radii of the ejecta components are determiend by
\begin{eqnarray}
	r_{\max,\rm w}(t)=r_{\max,\rm w}(0)+ v_{\mathrm{max,w}}t,\\
	r_{\max,\rm d}(t)=r_{\max,\rm d}(0)+ v_{\mathrm{max,d}}t,
\end{eqnarray}
where $v_{\mathrm{max,w}}$ and $v_{\mathrm{max,d}}$ are the corresponding maximum velocities. 

\begin{table}[htpb]
	\centering
	\begin{tabular}{ccccccc}
		\hline \hline
		Model &  A   & B  & C  & D  & E  & F  \\
		\hline
		$L_{\mathrm{w,i}}/\mathrm{erg\,s^{-1}}$ &&& $10^{48} $&&\\ 
		$ t_{\mathrm{md}}/\mathrm{s}$   &&& $5\times10^3  $ &&\\
		$\kappa_{\mathrm{w}}/\mathrm{cm^{2}\,g^{-1}}$  & & &1&& \\
		$\kappa_{\mathrm{d,p}}/\mathrm{cm^{2}\,g^{-1}}$  & && 1  &&  \\
			$\kappa_{\mathrm{d,e}}/\mathrm{cm^{2}\,g^{-1}}$  & && 10  &&  \\
		$\kappa_{\mathrm{x}}/\mathrm{cm^{2}\,g^{-1}}$  && &$7\times10^3 $&&\\
		$f$  & $0.01$ &  $0.1$&  $0.01$&  $0.01$&  $0.01$&  $0.01$\\
		$ M_{\mathrm{ej,w}}/ 0.01M_{\odot} $ &   $1$&  $1$&   $0.25$&  $1$&  $1$&  $1$\\
		$ M_{\mathrm{ej,d}}/0.01M_{\odot} $ &   $1 $ &  $1 $&   $0.1 $&   $0.25 $&   $1 $&   $1 $\\
		$ v_{\mathrm{max,d}}/c $ &   $ 0.3 $&   $ 0.3 $&   $ 0.3 $&   $ 0.3 $&   $ 0.2 $&  $ 0.4 $\\
        $ v_{\mathrm{max,w}}/c $ &   &&   $ 0.1 $& &\\		
		\hline    
	\end{tabular}
	\caption{Model parameters}
	\label{table1}
\end{table}

\subsection{Magnetar wind}
The remnant NS of a merger event is very likely to be in differential rotation initially, which can lead to the amplification of the magnetic fields of the NS through dynamo mechanisms (e.g. \citealp{Duncan1992,Price2006,Cheng2014}). As the NS become a magnetar, a powerful relativistic wind can be driven with a luminosity of
\begin{equation}
	L_{\mathrm{w,i}} =9.6\times 10^{48}B_{\mathrm{p,}15}^{2}R_{\mathrm{s,}6}^{6}P_{\rm i,-3}^{-4}\mathrm{erg\ s}^{-1},
\end{equation}
where $ B_{\rm p} $, $ R_{\mathrm{s,}} $ and  $ P_{\mathrm{ i}} $ are the polar magnetic filed strength, radius, and spin period of the magnetar. Here the convention $ Q_{x}/10^{x} $ is adopted in cgs units. The reference values of the parameters in the above expression  are adopted by according to the inference from the shallow-decay and plateau afterglows of GRBs by ascribing these afterglow features to the consequence of a magnetar engine \cite{Yu2010,Lyons2010,Lu2014,Lu2015}. Nevertheless, the fitting of the mergernova AT2017gfo, which
is in counterpart with the GW170817/GRB 170817A event, indicates that the magnetic filed
of the remnant NS on a timescale of days could be not much higher than $10^{11}$ G \citep{Yu2018,Li2018},  if such a NS indeed played a role in there as supposed. By
combining with these two different observational constraints, Yu et al. (2018) suggested that
the ultra-high magnetic field of the remnant magnetar could be suppressed drastically by
some unclear mechanisms (e.g., hidden by fallback material) at a time of a few hundreds
to thousand of seconds \cite{Yu2018}. This assumption could also be supported by the discovery of
an extremely steep decay following a plateau in some GRB afterglows. In any case, for simplicity, here we would not consider this potential complex in our calculation and the temporal evolution of the wind luminosity is adopted to be as usual as
\begin{equation}
L_{\mathrm{w}}\left(t\right)=L_{\mathrm{w,i}} \left(1+\frac{t}{t_{\mathrm{md}}}\right)^{-2},
\end{equation}
where $ t_{\mathrm{md}}=2\times 10^3R_{\mathrm{s,}6}^{-6}B_{\mathrm{p,}15}^{-2}P_{\mathrm{ i,-3}}^{2}\,\mathrm{s}$ is the spin-down timescale.

\subsection{Shock dynamics and breakout emission}
As the powerful magnetar wind collides with the merger ejecta, the wind can be shocked by the TS to generate high energy emission, which is absorbed in part by the merger ejecta. Meanwhile, the FS propagates into and finally crosses the ejecta and its the dynamical evolution can be determined by \cite{Li2016}
\begin{equation}
	\frac{dv_{\mathrm{sh}}}{dt} =\frac{1}{M_{\mathrm{sh}}v_{\mathrm{sh}}}\left[L_{\mathrm{w}}-L_{\mathrm{tot}}-\frac{1}{2}\left(v^2_{\mathrm{sh}}-v_{\mathrm{ej}}^2\right)\frac{dM_{\mathrm{sh}}}{dt}-\frac{dU}{dt}\right],
\end{equation}
where $ v_{\mathrm{sh}} $ is the shock velocity of the shock front, $U$ is the total internal energy of the shocked region, and $L_{\mathrm{tot}}$ is the total luminosity of the thermal emission of merger ejecta. The increase of the swept-up mass $M_{\mathrm{sh}} $ can be calculated by 
\begin{equation}
	\frac{d M_{\mathrm{sh}}}{d t } = 4\pi r_{\mathrm{sh}}^2\rho_{\mathrm{ej}}\left(v_{\mathrm{sh}}-v_{\mathrm{ej}}\right),
\end{equation}
where $r_{\rm sh}$ is the radius of the shock front, $v_{\mathrm{ej}}$ is the velocity of the ejecta material in front of the shock, and the mass density $\rho_{\rm ej}$ is given by Eq. (\ref{rhor}). Meanwhile, the thermal energy deposited in the shocked material is given by
\begin{equation}
	H_{\mathrm{sh}} =\frac{1}{2}\left(v_{\mathrm{sh}}-v_{\mathrm{ej}}\right)^2\frac{d M_{\mathrm{sh}}}{d t } .
\end{equation}
Here, since the merger ejecta is anisotropic, the above equations should be solved separately for different directions. After the determination of $H_{\rm sh}(t)$, the evolution of the internal energy accumulated by the shock can be derived from
\begin{equation}
	\frac{d U_{\mathrm{sh}}}{dt} = H_{\mathrm{sh}}- P_{\mathrm{sh}}\frac{d V_{\mathrm{sh}} }{dt} -L_{\mathrm{sh}},
\end{equation}
where  $ V_{\mathrm{sh}}$ and $ P_{\mathrm{sh}} =U_{\mathrm{sh}}/\left(3 V_{\mathrm{sh}}\right) $  are the volume and pressure of the shock heating region, respectively, and $L_{\rm sh}$ is the bolometric luminosity of the SBO emission. Meanwhile, the evolution of the internal energy excluding the shock-accumulated part, $U_{\mathrm{ej}}=U-U_{\mathrm{sh}}$, can be written as (ignoring the radioactivity energy)
\begin{equation}	
	\frac{dU_{\mathrm{ej}}}{d t} =  L_{\mathrm{w}}-P_{\mathrm{ej}}\frac{d V_{\mathrm{ej}} }{dt}- L_{\mathrm{mn}},
\end{equation}
where $V_{\mathrm{ej}}$ and $ P_{\mathrm{ej}} =U_{\mathrm{ej}}/\left(3 V_{\mathrm{ej}}\right) $ are the volume and the pressure of the ejecta behind the shock, respectively, and $L_{\rm mn}$ is the bolometric luminosity of the mergernova emission. 

Following the approximate one-zone diffusion model in Reference \cite{kasen2010}, the bolometric luminosity of the SBO emission can be estimated by\footnote{Here the shocked energy is considered to distribute approximately uniformly behind the shock, which is different from \cite{Li2016} where the shocked energy is assumed to be concentrated within a thin shell immediately behind the shock front.}
\begin{equation}
	L_{\mathrm{sh}}\approx\frac{ r_{\mathrm{max,d}}^2 U_{\mathrm{sh}}c} {r_{\mathrm{max,d}}^3-r_{\mathrm{sbo}}^3}\left[\frac{1-e^{-\tau_{\mathrm{sbo}}}}{\tau_{\mathrm{sbo}}}\right],
\end{equation}
where $r_{\mathrm{sbo}}$ is the radius at which SBO occurs and $\tau_{\mathrm{sbo}}$ is the optical depth of the shell between the radii $r_{\mathrm{sbo}}$ and $r_{\mathrm{max,d}}$. Above $r_{\mathrm{sbo}}$, photons can escape from the merger ejecta, because the photon diffusion time in the outside shell is equal to the dynamical time, i.e.,  
\begin{equation}
	\tau_{\rm sbo}=\int_{r_{\rm sbo}}^{r_{\max,\rm d}}\kappa\rho dr={ct\over r_{\max,\rm d}-r_{\rm sbo}},
\end{equation}
from which the radius $r_{\rm sbo}$ can be derived, where $\kappa$ is the opacity. When $r_{\rm sbo}\ll r_{\max,\rm d}$, the above equation returns to its usual form as $\tau_{\rm sbo}=c/v_{\rm max,d}$. Meanwhile, the bolometric luminosity of the mergernova emission can be given by
\begin{equation}
	L_{\mathrm{mn}}	\approx\frac{ U_{\mathrm{ej}}c} {r_{\mathrm{max,d}}}\left[\frac{1-e^{-\left(\tau_{\mathrm{sh}}+\tau_{\mathrm{un}}\right)}}{\tau_{\mathrm{sh}}+\tau_{\mathrm{un}}}\right],
\end{equation}
where $ \tau_{\mathrm{sh}} \approx \kappa M_{\mathrm{sh}}/4\pi r_{\mathrm{sh}}^2 $ and $ \tau_{\mathrm{un}} =\int_{r_{\mathrm{sh}}}^{r_{\mathrm{max,d}}}\kappa \rho dr $ are the optical depths of shocked and unshocked ejecta. In our calculations, the opacity $\kappa$ is adopted to be \cite{chen2021,Ren2022}
\begin{equation}
\kappa=	\kappa_{\mathrm{x}}\left(\frac{3.92 k T_{\mathrm{int}}}{1\mathrm{keV}}\right)^{-1}
\end{equation}
for $k T_{\mathrm{int}}\ge 0.3\mathrm{keV}$ and  
\begin{equation}
	\kappa=\begin{cases}
		\kappa_{\mathrm{w}}, & {\rm for~the~wind~component},\\  
		\kappa_{\mathrm{d,p}}, &  {\rm for~the~dynamical~component~of~} \theta\le\pi/4,\\
		\kappa_{\mathrm{d,e}}, & {\rm for~the~dynamical~component~of~} \theta>\pi/4,
	\end{cases}
\end{equation}
for $k T_{\mathrm{int}}< 0.3\mathrm{keV}$, where $T_{\mathrm{int}}\approx \left(cU_{\rm }/4V_{\rm }\sigma\right)^{1/4}$ is the average internal temperature of the related region, $k$ is the Boltzmann constant, $ \sigma $ is the Stefan-Boltzmann constant, and the values of parameters $\kappa_{\mathrm{x}}$, $\kappa_{\mathrm{w}}$, $\kappa_{\mathrm{d,p}}$ and $\kappa_{\mathrm{d,e}}$ are listed in Table \ref{table1}.  Here, for the dynamical ejecta, two different opacity values are simply taken for $\theta\le\pi/4$ and $\theta>\pi/4$, in order to reflect the angluar-dependence of the electron faction, which is result of the anisotropic neutrino irradiation from the remnant NS and the disk.

Finally, as photons diffuse and are thermalized continuously, a black-body emission appears at a photosphere of radius $r_{\rm ph}$ with an effective temperature of
\begin{equation}
	T_{\mathrm{eff}} =\left(\frac{L_{\mathrm{tot}}}{4 \pi \sigma r_{\mathrm{ph}}^2}\right)^{1/4} ,
\end{equation}
where $ L_{\mathrm{tot}}= L_{\mathrm{sh}}+	L_{\mathrm{mn}} $ and $  r_{\mathrm{ph}}\approx r_{\mathrm{max,d}} $ is taken in our calculation. Then, the emergent net flux at a frequency $ \nu $ can be given by using the black-body spectrum as
\begin{equation}
	F_{\nu} =\frac{2\pi h \nu^3}{c^2}\frac{1}{\mathrm{exp}\left(h\nu/ kT_{\mathrm{eff}}\right)-1},\label{Fnu}
\end{equation}
where $ h $ is the Planck constant.

%%%%%%%%%%%%%%%%%%%%%%%%%%%%%%%%%%%%%%%%%%
\subsection{Integration over the emission surface}\label{Int}
\begin{figure*}[htpb]
	\centering
	\includegraphics[width = 0.7\linewidth ]{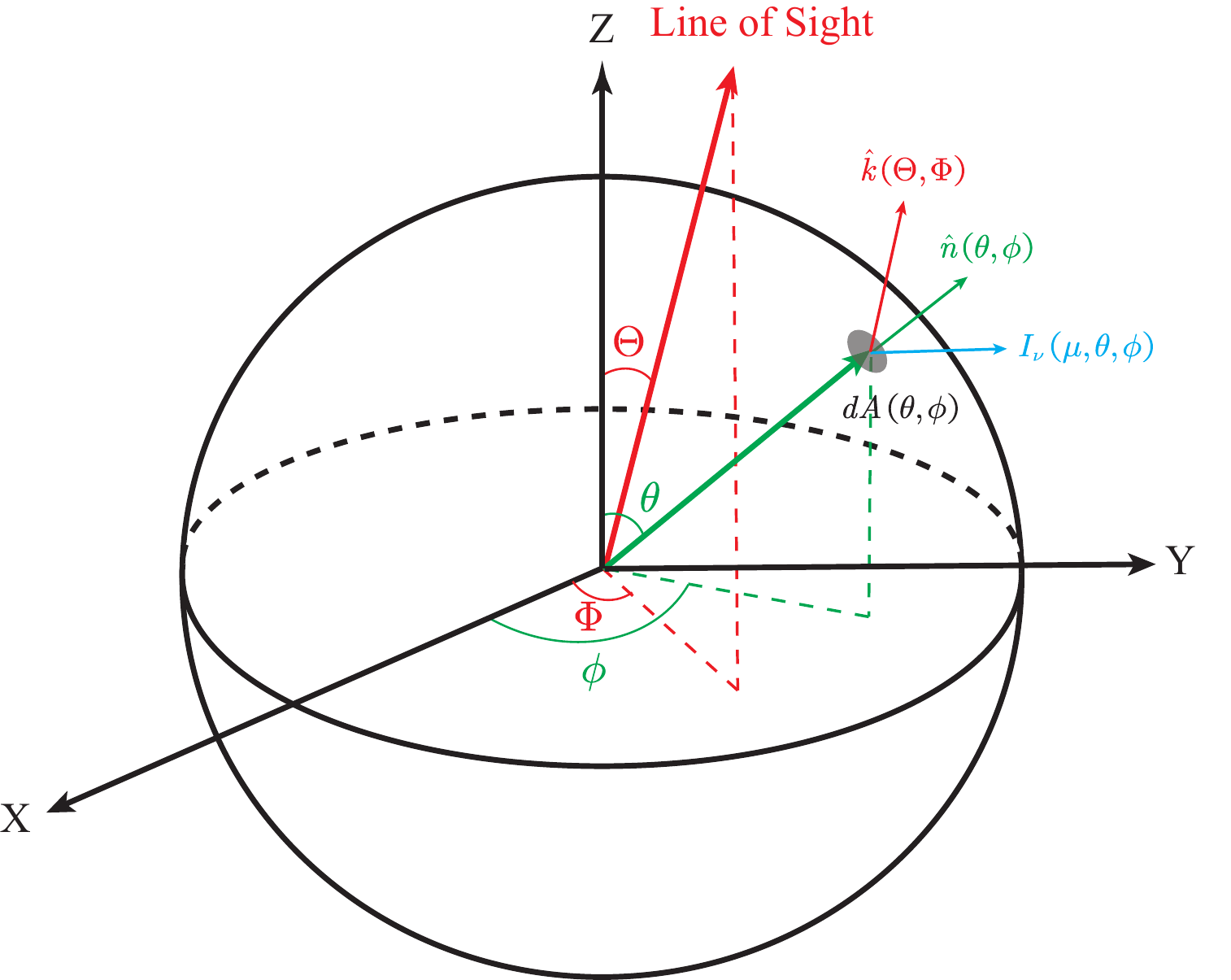}
	\caption{The sketch of the coordinate system we use, the angles $\Theta$ and $\Phi$ are the represent the direction of line of sight, the angles $\theta$ and $\phi$ represent the location of a point in local structure.}
	\label{fg0}
\end{figure*}
As illustrated in Fig. \ref{fg0}, we adopt coordinates with $z-$axis along the symmetric axis of the system and denote the zenith and azimuth angles by $\theta$ and $\phi$. Then, the specific luminosity for an observer on the line of sight (LOS) of $(\Theta,\Phi)$ can be obtained by integrating the specific intensity over the visible photosphere as
\begin{equation}\label{sL}
	L_{\nu}\left(\Theta,\Phi\right) = 4\pi r_{\rm ph}^2 \int_{\mu\ge 0}^{} I_{\nu}\left(\mu,\theta,\phi\right)\mu	d\Omega \left(\theta,\phi\right),
\end{equation}
which is viable for an unresolved emitting source \cite{Rybicki1969,Collins1973,Kandel1973,Rybicki1983}, where $\mu =\hat{n}\left(\theta,\phi\right)\cdot \hat{k}\left(\Theta,\Phi\right)$ with $\hat{n}$ and $\hat{k}$ being the unit vectors of the normal of the differential solid angle and the direction of the LOS, respectively. For an ideal black-body surface, the angular distribution of emergent intensities is uniform and, thus, the value of $I_{\nu}$ can in principle be derived from Eq. (\ref{Fnu}) to $I_{\nu}\left(\mu,\theta,\phi\right) = F_{\nu}\left(\theta,\phi\right) /\pi$. However, as a more realistic consideration, the hot, expanding ejecta should contain an abundance of free electrons, especially, in the fastest head of dynamical ejecta. In other words, the atmosphere above the photosphere can deviate from the thermal equilibrium and become scattering dominated. In this case, the limb darkening effect needs to be taken into account, which is widely discussed in the radiation transfer of stellar atmosphere \cite{Hopf1930,Chandrasekhar1960,Mihalas1978}.
Here we simply consider pure isotropic scattering, which leads the emergent intensities to be different for different directions as \cite{Bronstein1929,Hopf1930,Chandrasekhar1960}  
\begin{equation}\label{sI}
	I_{\nu}\left(\mu,\theta,\phi\right) =\frac{5}{4} \mathcal{H}\left(\mu\right)F_{\nu}\left(\theta,\phi\right) .
\end{equation}
where\footnote{An exact expression of $ \mathcal{H}\left(\mu\right)$ had been given by \citeauthor{Mihalas1978} (\citeyear{Mihalas1978}) as
	\begin{equation}
		\mathcal{H}\left(\mu\right)=\frac{1}{\left(1+\mu\right)^{1/2}}\mathrm{exp}\left[\frac{1}{\pi}\int_{0}^{\pi/2}\frac{\theta \tan^{-1}\left(\mu\tan\theta\right)}{1-\theta\cot\theta}\right].
\end{equation}}
\begin{equation}
	\mathcal{H}\left(\mu\right)=\frac{I_{\nu}\left(\tau=0,\mu\right)}{I_{\nu}\left(\tau=0,\mu=1\right)}=\frac{3}{5}\left[(\mu+q\left(\tau=0\right)\right]
\end{equation}
for $\mathcal{H}\left(\mu=1\right)=1$ and $ q\left(\tau\right) $ is the Hopf function. This function implies that the intensity emerging from the limb is about $ 40\%$ of that from the center of the emission source, when the Eddington approximation is taken $ q\left(\tau=0\right)=2/3 $, which is in fair agreement with the observations of the Sun.

\begin{figure*}[htpb]
	\centering
	\includegraphics[width = 1\linewidth ]{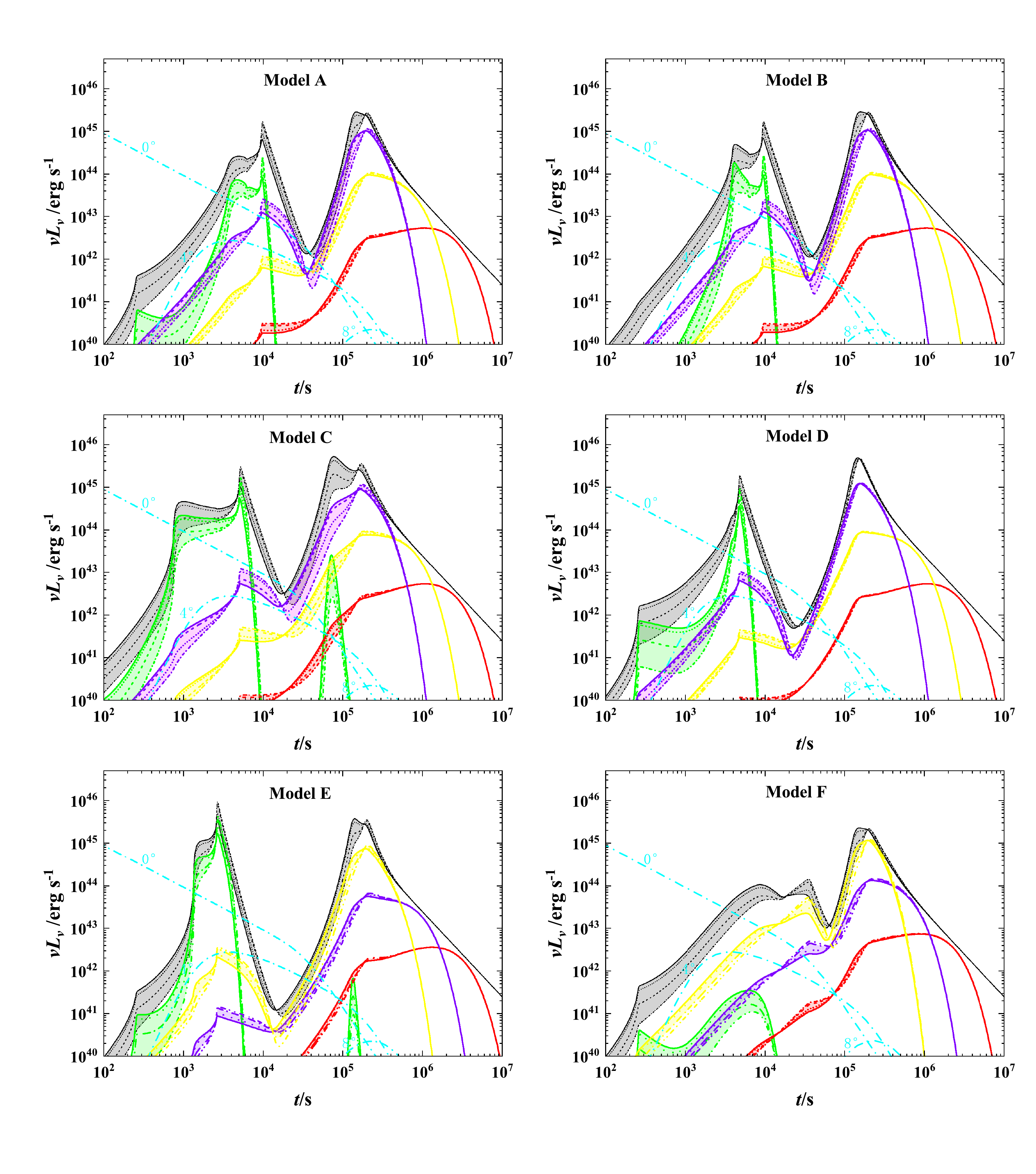}
	\caption{Mult-band light curves of the magnetar-driven SBO (i.e., the first bump at around $10^{3-4}$ s) and mergernova (i.e., the second bump at around $10^{5-6}$ s) emission from an anisotropic merger ejecta for parameters listed in Table \ref{table1}, including 0.1 keV (Green), 6 keV (Purple), 2 eV (Yellow), 0.5 eV (Red) and the bolometric ones (black). For a comparison, the GRB X-ray ($ 0.1 \,\mathrm{KeV} $) afterglow is also presented by the cyan dotted-dashed lines for three viewing angles of $ 0^{\circ} $, $ 4^{\circ} $ and $ 8^{\circ} $, which are yielded from the \textit{afterglowpy} \cite{Ryan2020} for a top-hat jet model with a half-opening angle of $ 3^{\circ} $, where the typical parameter values as inferred from the GRB 170817A observations are taken as $ E_{\mathrm{iso}}=10^{51}\, \mathrm{erg} $, $ n=10^{-3}\, \mathrm{cm}^{-3} $, $ p=2.3 $, $ \epsilon_{e}=0.1 $, and $\epsilon_{\mathrm{B}}=10^{-4} $. Due to the limb darkening effect, the SBO emission can slightly vary with the observational directions, as labeled by the solid ($\Theta=0^{\circ}$), dotted ($ 30^{\circ} $), dashed ($ 60^{\circ} $), and dashed–dotted ($ 90^{\circ} $) lines. }
	\label{fg2}
\end{figure*}

\begin{figure*}[htpb]
	\centering
	\includegraphics[width = 1\linewidth ]{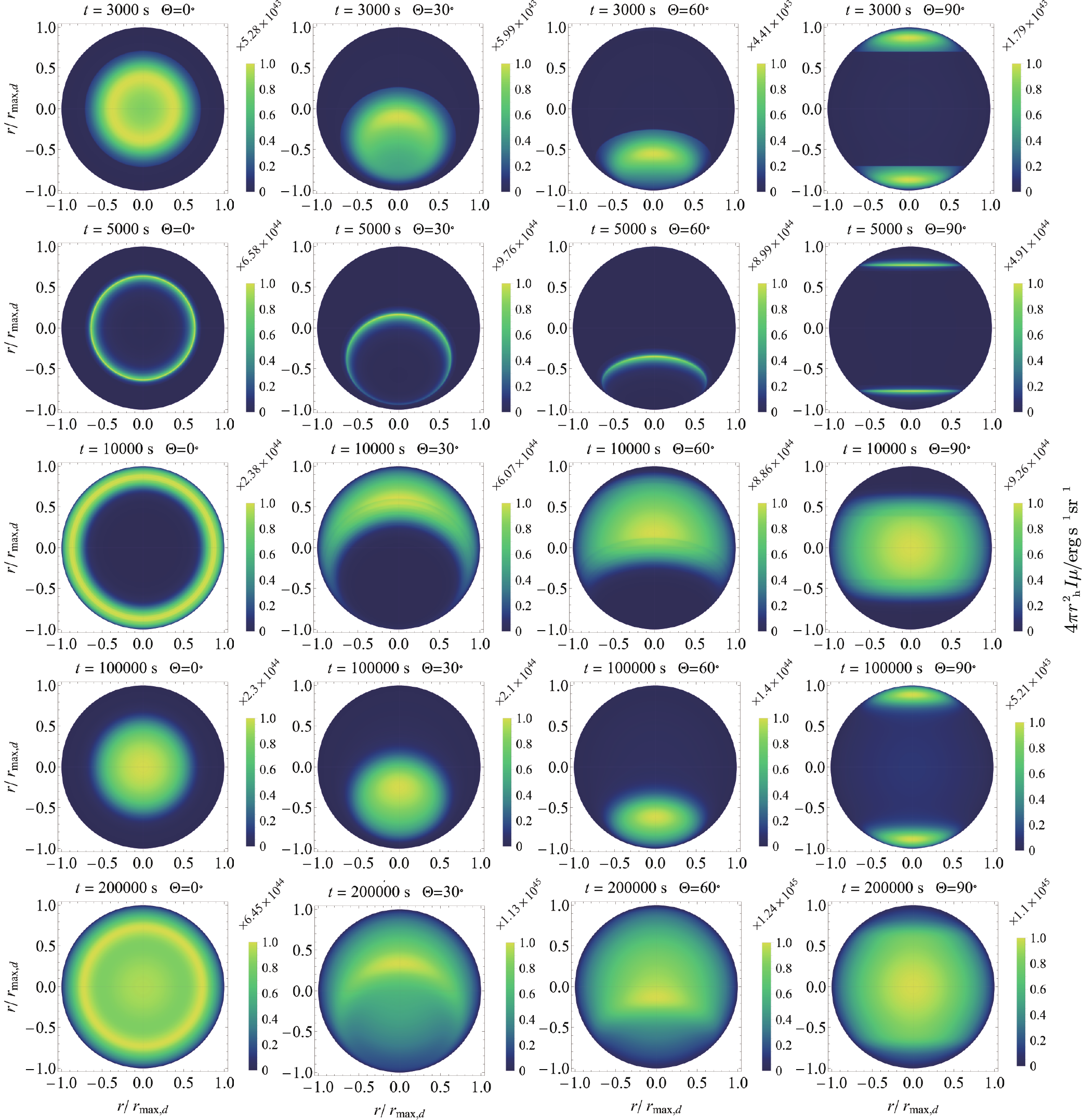}
	\caption{The intensity distribution on the emission surface for different times (rows) and different viewing angles (columns), for the Model A parameter values.}\label{LOSdepend}
\end{figure*}

%%%%%%%%%%%%%%%%%%%%%%%%%%%%%%%%%%%%%%%%%%
\section{Results}
For the parameter values listed in Table \ref{table1}, we calculate the light curves of the SBO and mergernova emission, as presented in Fig. \ref{fg2}. First of all, the SBO emission is mainly in the UV and soft X-ray bands, while the mergernova emission is in optical, just as found in \cite{Li2016}. In comparison with the GRB afterglow emission, the SBO emission is most likely to be detected in soft X-rays, especially, when the GRB jet is deviate from the LOS. 

The anisotropic structure of the merger ejecta can broaden the bolometric light curve of the SBO emission. This is because the SBO occurs at different times for different directions and thus it takes a relatively long period to observe the total SBO emission. In more detail, the harder emission can emerge earlier than the softer emission, since the former corresponds to a smaller SBO radius due to a thin merger ejecta in the polar direction while the later comes from the thick equatorial ejecta. Such a shift of the highlight on the emission surface is clearly shown in Fig. \ref{LOSdepend}.  In more detail, the comparison between the different panels shows that the profile of the light curves, in particular, of the increasing phase is highly dependent on the structure parameters of the ejecta, which thus enables us to probe the ejecta structure by observing this SBO emission.

In addition, the SBO light curves show a slight dependence of the viewing angle. This a result of the limb darkening effect, which leads to the on-axis observation on an emitting element can receive more intensity than the off-axis observation. Therefore, when the highlight point moves on the emission surface, observers on different directions would detect different light curves, as also shown in Fig. \ref{LOSdepend}. Here, it is worth mentioning that this viewing angle-dependence of the SBO emission would be likely to cause a significant difference in the polarization of the emission between different observational directions, although its influence on the light curve profile is unremarkable.

%%%%%%%%%%%%%%%%%%%%%%%%%%%%%%%%%%%%%%%%%%
\section{Conclusions and Discussion}
In this paper, we investigate the influence of the anisotropic structure of merger ejecta on the SBO emission driven by a magnetar wind. In view of the possible suppression of the magnetic field of the magnetar at late time, the early SBO emission (in particular, in the soft X-ray band) can provide an important observational signature for the existence of a remnant magnetar. 

It is found that, due to the different density distribution in different directions, the occurring time of the SBO can vary with the angle relative to the symmetric axis of the system. This leads to the broadening of the bolometric light curve of the SBO emission, which makes it more possible to be detected. Furthermore, in future, as some detailed observations can be achieved (e.g., by the Einstein Probe \cite{Yuan2022}), we can even use the SBO light curve to probe the structure of the merger ejecta, where of course a more realistic structure description must be involved for elaborate modelings. In addition, we also find the observational SBO emission can be slightly dependent on the viewing angle. This is because the anisotropic structure of the merger ejecta leads to a highlight point moving from the polar to the equatorial direction on the emission surface. This could simultaneously lead to a significant variation in the polarization of the emission for different observational directions, which is worth to be investigated carefully in future.

Recently, some fast X-ray transients have been detected and suspected to be powered by a spinning-down magnetar \cite{Xue2019,Lin2022}, which could be directly contributed by the dissipating magnetic wind if it is not trapped by thick material \cite{Zhang2013}. Alternatively, as presented in this paper, even if the magentar wind is surrounded by a merger ejecta, the SBO driven by the magnetar wind could still generate an X-ray transient. In contrast to the highly-beamed GRB emission, this SBO emission can in principle be detected at arbitrary directions, which therefore provides a valuable signal for test the existence of the remnant magnetar. Additionally, in view of the similarity of the mergernovae and superluminous supernovae, such a SBO signal can also be expected to appear in these supernova emission, probably with a relatively longer timescale \cite{Liu2021,Zhang2022}.

\section*{Acknowledgments}
This work is supported by the National Key R\&D Program of China (2021YFA0718500), the National SKA Program of China (2020SKA0120300), and the National Natural Science Foundation of China (grant No. 11833003).
\vspace{6pt} 

%%%%%%%%%%%%%%%%%%%%%%%%%%%%%%%%%%%%%%%%%%
%% Optiona

%%%%%%%%%%%%%%%%%%%%%%%%%%%%%%%%%%%%%%%%%%
\begin{adjustwidth}{-\extralength}{0cm}
	\printendnotes[custom] % Un-comment to print a list of endnotes
	
	\reftitle{References}
	%=====================================
	% References, variant A: external bibliography
	%=====================================

\end{adjustwidth}

\end{document}